\documentclass[twocolumn]{aa}
\usepackage{graphicx}
\usepackage{txfonts}
\usepackage{natbib}
\bibpunct{(}{)}{;}{a}{}{,}

\begin{document}

\title{ The HARPS search for southern extra-solar planets\thanks{Based
    on observations made with the HARPS instrument on the ESO 3.6 m
    telescope at La Silla Observatory under the GTO programme ID
    072.C-0488. 
}  }
\subtitle{XIII. A planetary system with 3 Super-Earths (4.2, 6.9, \&
  9.2\,M$_{\oplus}$)}

\author{
M.~Mayor\inst{1}
\and S.~Udry\inst{1}
\and C.~Lovis\inst{1}
\and F.~Pepe\inst{1}
\and D.~Queloz\inst{1}
\and W.~Benz\inst{2}
\and J.-L.~Bertaux\inst{3}
\and F.~Bouchy\inst{4}
\and C.~Mordasini\inst{2}
\and D.~Segransan\inst{1}
}

\offprints{M. Mayor}

\institute{ 
  Observatoire de Gen\`eve, Universit\'e de Gen\`eve, 51 ch.
  des Maillettes, 1290 Sauverny,  Switzerland
  \email{michel.mayor@obs.unige.ch}
  \and Physikalisches Institut, Universit\"at Bern, Silderstrasse 5, 
CH-3012 Bern, Switzerland
  \and Service d'A\'eronomie du CNRS/IPSL, Universit\'e de Versailles
  Saint-Quentin, BP3, 91371 Verri\`eres-le-Buisson, France 
  \and Institut d'Astrophysique de Paris, CNRS, Universit\'e Pierre et
  Marie Curie, 98bis Bd Arago, 75014 Paris, France 
}

\date{Received ; accepted To be inserted later}

\abstract{This paper reports on the detection of a planetary system
  with three Super-Earths orbiting HD\,40307. HD\,40307 is a K2\,V
  metal-deficient star at a distance of only 13\,parsec, part of the
  HARPS GTO high-precision planet-search programme. The three planets
  on circular orbits have very low minimum masses of respectively 4.2, 6.9 and
  9.2 Earth masses and periods of 4.3, 9.6 and 20.5 days.  The planet
  with the shortest period is the lightest planet detected to-date
  orbiting a main sequence star. The detection of the correspondingly
  low amplitudes of the induced radial-velocity variations is
  completely secured by the 135 very high-quality HARPS observations
  illustrated by the radial-velocity residuals around the 3-Keplerian solution of
  only 0.85\,ms$^{-1}$.  Activity and bisector indicators exclude any
  significant perturbations of stellar intrinsic origin, which
  supports the planetary interpretation.  Contrary to most planet-host
  stars, HD\,40307 has a marked sub-solar metallicity
  ([Fe/H]\,=\,$-$0.31), further supporting the already raised
  possibility that the occurrence of very light planets might show a 
  different dependence on host 
  star's metallicity compared to the population of gas giant
  planets.  In addition to the 3 planets close to the
  central star, a small drift of the radial-velocity residuals
  reveals the presence of another companion in the system the nature
  of which is still unknown.

 }

\maketitle

\section{Introduction and context}

The planet-search programme conducted at high precision with the HARPS
spectrograph on the ESO 3.6-m telescope at La Silla aims at the
detection of very low-mass planets in a sample of solar-type stars
already screened for giant planets at a lower precision with
CORALIE on the 1.2-m Swiss telescope on the same site. About 50\,\% of
the HARPS GTO time is dedicated to this survey. After 4.5 year of the
programme, we are starting to see a large population of Neptune-mass
and super-Earth planets emerging from the data, including the system
presented here.

Several reasons motivate our interest to search for very low-mass
planets, with masses in the range of the Neptunes or the so-called
Super-Earths ($\sim$\,2\,M$_{\oplus} \leq m_2\sin{i} \leq 10$\,M$_{\oplus}$).

i) Over the past decade, several statistical distributions of the
orbital elements of gaseous giant planets have emerged from the nearly
300 detected planetary systems \citep[see
  e.g.][]{Udry-2007:b,Marcy-2005}. These statistical properties provide
constraints to complex physical scenarios of planetary system
formation. One of the most obvious example of that dialogue between
planetary formation theory and observations is illustrated by the
comparison of the planetary mass vs semi-major axis ($m_2-a$) diagram
\citep{Ida-2004:a,Mordasini-2008}. Comparison can be made for specific 
categories of host stars by selecting different primary masses ($m_1$) 
or metallicities ([Fe/H]). In all cases, global features of planet formation
directly affect the overall topology of the ($m_2-a$) diagram. In particular, 
the location of the large population of very low-mass planets predicted by 
the models  \citep{Mordasini-2008,Ida-2008} depends upon the extend 
of migration undergone by the planets during their formation. The detection 
of a large sample of planets with masses less than roughly 25\,M$_{\oplus}$
at relatively close distances is therefore an important indicator of the efficiency
of type I migration (assuming that the planets are not too close so that evaporation 
can be neglected). Despite the still
very limited number of planets detected in the range of Neptune
masses, already a few interesting characteristics are emerging
\citep{Mayor-2008}:
 
- The distribution of planetary masses appears as bimodal. A new
population of light planets, although more difficult to detect, is
differentiating itself from the distribution of giant-planet masses.
The Neptune- and super-Earth mass distribution seems not to be the
extrapolation towards lower masses of the distribution for gaseous
giant planets.

- The very strong correlation observed between the host star
metallicity and occurrence frequency of giant planets 
\citep{Santos-2001:a,Santos-2004:b,Fischer-2005} seems to be vanishing
or at least to be reduced \citep{Udry-2006}.

- Neptune-mass planets and Super-Earths are found most of the time in
multiplanetary systems ($>$\,80\,\% of the known candidates).

ii) Simulations of planetary formation do not only provide the
statistical distributions of masses and semi-major axes. For every
planet, we have, in addition, a prediction of its internal
structure. The internal composition of the planet at the end of the
formation/migration process carries a fossil signature of the system
history. The end-state diversity is broad: rocky planets, icy planets,
ocean planets, evaporated gaseous giant planets, or possibly objects
with variable percentages of these ingredients.  The predicted
distributions of the planetary internal composition, as a function of
the different significant parameters ($m_1$, $m_2$, $a$, [Fe/H]) can
be observed in the corresponding radius-mass ($m_2 - R$) diagrams.
These predicted ($m_2 - R$) distributions can then be compared to the
observed distributions derived from combined radial-velocity and
transit searches.  The coming soon results from space missions
searching for planetary transits combined with ground-based,
high-precision, radial-velocity follow-up, will provide rich
observational constraints to the planetary formation theory via the
$m_2-R$ distributions of low-mass planets.
 
iii) In a more distant future, space missions will be developed to
search for life signatures in the
atmosphere of terrestrial-type planets. Before the detailed design of
such ambitious missions, it would be wise to have first insights in
the occurrence frequency of terrestrial planets and on the statistical
properties of their orbits. Still more valuable is the detection of
planets in the habitable zone of our closest neighbours, initiating
the preparation of an "input catalogue" for these future missions.

Already a few planets have been detected with masses between 3 and
10\,M$_{\oplus}$.  In 1992, from precise timing, \citep{Wolszczan-92}
have discovered two planets with masses of 2.8 and 3.4\,M$_{\oplus}$
orbiting the pulsar PSR1257+12 on almost perfect circular orbits.
Microlensing technique has demonstrated as well its potential to detect
low-mass planets. Two planets with masses possibly in the range of
super-Earths have been announced: a planet of about 5.5\,M$_{\oplus}$
orbiting a low-mass star \citep{Beaulieu-2006} and a still less massive
object, of only 3.3\,M$_{\oplus}$, probably gravitationally bound to a
brown dwarf \citep{Bennett-2008}.  Doppler spectroscopy also revealed
quite a few planets with $m_2\sin{i}$ less than 10\,M$_{\oplus}$:
GJ\,876\,d, $m_2\sin{i}=5.9$\,M$_{\oplus}$ \citep{Rivera-2005};
GJ\,581\,c and d, with $m_2\sin{i}$ of 5.1 and 8.2\,M$_{\oplus}$,
respectively \citep{Udry-2007:a}; HD\,181433\,b with
$m_2\sin{i}=7.5$\,M$_{\oplus}$ \citep{Bouchy-2008}. GJ\,876 as well as
GJ\,581 are both stars at the bottom of the main sequence with
spectral types M4\,V and M3\,V, respectively. HD\,181433 and HD\,40307
are on the other hand both K dwarfs.  The detection of planetary
systems around bright stars by Doppler spectroscopy is very important
because it concerns mostly nearby stars for which we can obtain
interesting further information (e.g. on their orbital, planetary and
stellar properties), opening the possibility of rich complementary
studies. This is even more true when the planet is transiting in front
of its parent star (see for illustration the wealth of studies concerning
GJ\,436).

In this paper, we characterize the orbits of three new super-Earth
planets on short-period trajectories orbiting the main sequence star HD\,40307.
Section\,2 will briefly describe the improvement of the HARPS
radial velocities in term of observational strategy and 
software developments. The stellar characteristics of HD\,40307 are
presented in Sect.\,3. Section\,4 deals with the derived orbital solution
while summary and conclusion are given in Sect.\,5.

\section{Challenges to high-precision radial velocities}

\subsection{Star sample and observational strategy} 

About 50\,\% of the HARPS GTO time (guaranteed time observation) is
devoted to the search for very low-mass planets orbiting G and K
dwarfs of the solar vicinity, in the southern sky. A companion
programme is focused on the same goal but for planets around M-dwarf
stars. HARPS is a vacuum high-resolution spectrograph 
fiber-fed by the ESO 3.6-m telescope at La Silla Observatory
\citep{Mayor-2003:a}. HD\,40307 is part of our G--K survey. The stars in
this programme were selected from the volume-limited sample (1650
stars) measured since 1998 with the CORALIE spectrograph on the EULER
telescope, also at La Silla \citep{Udry-2000:a}.  In a first step we have
selected some 400 G and K dwarfs with $v\sin{i}$ less than 3\,kms$^{-1}$ and
removing the known spectroscopic binaries. In a second step, using the
high quality HARPS spectra, we have concentrated our efforts on the
less active stars of the sample ($\log R^{\prime}_{HK}<-4.8$).  The
remaining $\sim$\,200 stars with rather low chromospheric activity
constitute the core of the programme to search for Neptune-mass and
super-Earth planets.

The planetary minimum mass estimated from Doppler measurements is directly
proportional to the amplitude of the reflex motion of the primary
star. The measure of very precise radial velocities requires that all
steps along the light path, from the star to the detector, are well
understood. If photon supply is no longer a problem (e.g. for bright
stars) there still remain two main limitations to the achieved
radial-velocity precision: intrinsic stellar variability and
spectrograph stability. The stellar noise groups error sources related
to stellar intrinsic phenomena acting on different time scales, from
minutes (p modes), to hours (granulation) and even days or weeks (activity). To
minimize their effects on the measured radial velocities we have to
adapt as much as possible our observational strategy to these
corresponding time scales, in such a way that averaged radial-velocity
values will be much less sensitive to the mentioned effects.  For
example, long integrations ($\sim$15~minutes) are sufficient to damp
the radial-velocity variations due to stellar oscillations well below
1\,ms$^{-1}$. To damp the granulation noise several measurements spread
over a few hours will probably be required. Test observations with
HARPS are ongoing to better characterize this point. We finally are
trying to avoid activity effects by selecting the less active stars,
as mentioned above.

\subsection{Precision improvements through software developments}

Recently the precision of the HARPS measurements has been notably
improved thanks to three major upgrades of the reduction software:

-- The precision of radial velocity measurements depends on many
factors, but in particular on the quality of the relation between
wavelengths and position on the CCD detector. This relation is
established before the beginning of each night by using the numerous
thorium lines of a thorium-argon hollow cathode lamp.  A global
reanalysis of thorium lines positions based on many thousand HARPS 
spectra  has allowed \citet{Lovis-2007:a} to
significantly improve the precision of thorium line
wavelengths. Immediately a drastic increase of the stability of the
HARPS dispersion relation has been observed.

-- Aging of the thorium-argon calibration lamp can be observed
producing a small wavelength shift of the emission lines due to the changing
pressure inside the lamp.  Such an effect is much larger for argon
lines than for thorium lines. This differential effect can be used to
correct for the lamp aging effect (Lovis, in prep).

-- The ADC (Atmospheric Dispersion Corrector) in the Cassegrain
adaptor is designed to maintain the stellar image at any wavelengths
precisely centered on the entrance of the optical fiber feeding the
spectrograph. But as the seeing is chromatically dependent on the
airmass, the relative amount of light entering in the optical fiber,
for the different grating orders, is depending on the airmass. The
used cross-correlation technique provides the mean
velocities of stellar lines order by order. Mean radial velocities
of the different orders are not exactly identical due to the random
distribution of the line blending, to the line-to-line dependence of 
the stellar convective blueshift, etc. To correct for
this secondary effect on the Doppler velocity, we must normalized the
flux received in the different orders to a given value and compensate
for the seeing chromatic dependence with airmass.

All the spectra obtained on the full span of HARPS measurements have
been reprocessed taking into account these new developments.  The net
effect is a clear improvement of the long term stability of the HARPS
measurements (better than a fraction of a m/s over 5 years) as well as
a significant decrease of the observed radial-velocity rms of the
stable stars.  The planetary mass estimated from Doppler measurements
is directly proportional to the amplitude of the stellar reflex
motion. Our progress to detect very low-mass planets is then directly
related to the progress done to improve the sensitivity and stability
of spectrograph. Ice giants and super-Earths induce RV amplitudes smaller 
than 3 m/s. This makes the quest for the highest possible RV precision
critical to detect this population.

\section{Stellar characteristics of HD\,40307}

The basic photometric (K2.5V, $V$\,=\,7.12, $B$\,$-$\,$V$\,=\,0.92)
and astrometric ($\pi$\,=\,77.95\,mas) properties of HD\,40307 were
taken from the Hipparcos catalogue \citep{Esa-1997}. They are recalled
in Table\,\ref{TableStar}, together with inferred quantities like the
absolute magnitude ($M_V$\,=\,6.63) and the stellar physical
characteristics derived from the HARPS spectra by \citet{Sousa-2008}.
These authors provide for the complete high-precision HARPS sample
(including HD\,40307) homogeneous estimates for the effective
temperature ($T_{\rm eff}$\,=\,4977\,$\pm$\,59\,K), metallicity
([Fe/H]\,=\,$-0.31$\,$\pm$\,0.03), and surface gravity
($\log{g}$\,=\,4.47\,$\pm$\,0.16) of the stars. 

Very interestingly HD\,40307 has a substellar metallicity with
[Fe/H]\,=\,$-0.31$ unlike most of gaseous giant-planet host stars
\citep{Santos-2004:b}.  According to simulations of planet formation
based on the core-accretion paradigm, moderate metal-deficiency does
not hamper the formation of low-mass planets
\citep{Ida-2004:b,Mordasini-2008}.  Taking into account its subsolar
metallicity, \citet{Sousa-2008} have also derived for the star a mass of
0.77\,M$_{\odot}$.  From the colour index, the derived effective
temperature, and the corresponding bolometric correction, we estimated
the star luminosity to be 0.23\,L$_\odot$.

HD\,40307 is among the least active stars of our sample with an
activity indicator $\log R^{\prime}_{HK}$ of -4.99. No significant
radial-velocity jitter is thus expected for the star.  From the activity
indicator we also derive a stellar rotation period
$P_{rot}$\,=\,$\sim$\,48\,days \cite[following ][]{Noyes-84}.

\begin{table}
\caption{Observed and inferred parameters of HD\,40307. Photometric
  and astrometric parameters were taken from the Hipparcos catalogue
  \citep{Esa-1997} and the stellar physical quantities from
  \citet{Sousa-2008}.}
\label{TableStar}
\centering
\begin{tabular}{l l c c}
\hline\hline
\multicolumn{2}{l}{\bf Parameter} &\hspace*{2mm} & \bf HD\,40307 \\
\hline
Sp & & & K2.5\,V  \\
$V$ & [mag] & & 7.17 \\
$B-V$ & [mag] & & 0.92  \\
$\pi$ & [mas] & & 77.95 $\pm$ 0.53 \\
$M_V$ & [mag] & & 6.63 \\
$T_{\mathrm{eff}}$ & [K] & & 4977 $\pm$ 59 \\
log $g$ & [cgs] & & 4.47 $\pm$ 0.16 \\
$\mathrm{[Fe/H]}$ & [dex] & & $-0.31$ $\pm$ 0.03 \\
$L$ & [$L_{\odot}$] & & 0.23  \\
$M_*$ & [$M_{\odot}$] & & 0.77 $\pm$ 0.05 \\
$v\sin{i}$ & [km s$^{-1}$] & & $<1$ \\
$\log R'_{\mathrm{HK}}$ & & & $-4.99$ \\
$P_{\mathrm{rot}}$($\log R'_{\mathrm{HK}}$) & [days] & & $\sim$\,48 \\
\hline
\end{tabular}
\end{table}

\section{The HD\,40307 planetary system}

\subsection{Radial velocity observations}

HD\,40307 has been measured with the HARPS spectrograph during 4.5
years.  135 measurements have been obtained.  During the last 3
seasons (128 measurements, span of 878 days), we have adopted a 
measurement strategy aiming at minimizing the effect of stellar acoustic 
modes. Despite the brightness of the star the integration time per epoch 
has always been set to 15 minutes. Over such an integration time, covering 
a few periods of the p-modes, the residual effect of these modes is estimated 
to be less than 20\,cms$^{-1}$. For most of the measurements, the photon 
noise after 15 minutes is typically of the order of 0.3\,ms$^{-1}$
(corresponding to S/N of the order of 150-200 at $\lambda=550$\,nm,
depending on seeing conditions). The mean uncertainty of the 128 
velocities obtained during the last 3 seasons is 0.32\,ms$^{-1}$.
As already mentioned, HD\,40307 presents a very low level of
chromospheric activity ($\log R^{\prime}_{HK}$\,=\,$-4.99$). The
radial velocity intrinsic variability including the noise due to
stellar granulation \citep{Kjeldsen-2005} and any effect related to the
stellar magnetic activity is not known but, due to the low level of
calcium reemission, this effect is by comparison with other similar
stars less than 1\,ms$^{-1}$.  The observed raw rms for the 135
measurements spread over 1628 days is 2.94\,m/s, much in excess of the
above-mentioned different sources of noise.

\begin{figure}[t!]
\centering
\includegraphics[width=0.9\hsize]{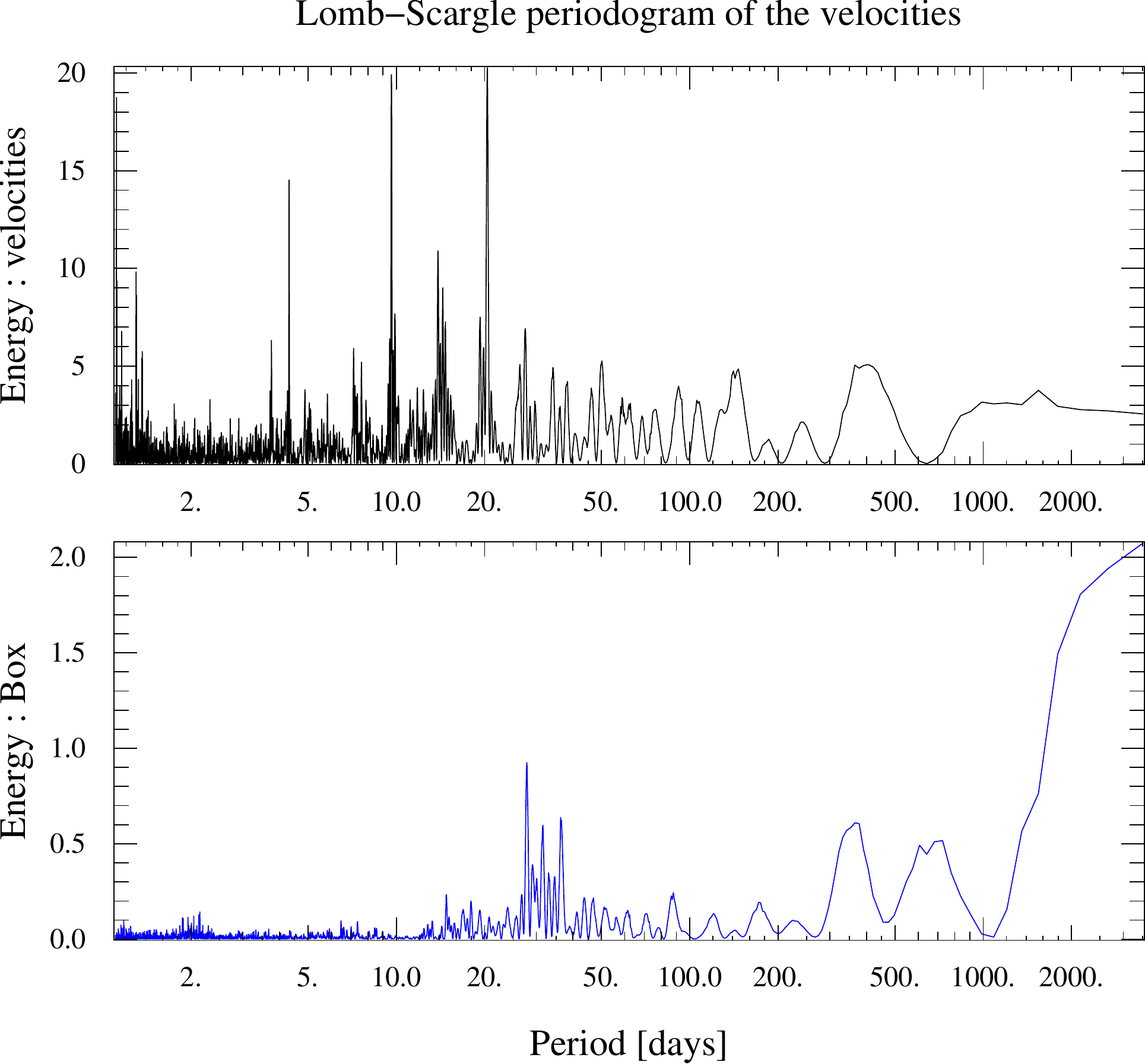}
\caption{ Lomb-Scargle periodogram (top) of the 135 HARPS radial
  velocities of HD\,40307. Clear peaks are found at frequencies
  corresponding to periods of 4.2, 9.6 and 20.5 days. The bottom
  panel presents the corresponding window function of the data.  }
\label{mayorfig1}
\end{figure}

End of 2006 we noticed that the observed radial-velocity variation
could be explained by the effect of 3 low-mass planets. A first
preliminary solution was already obtained. But due to the complicated
pattern of the radial velocities, the number of parameters required
for a 3-planet fit (16 free-parameters) and the low velocity
semi-amplitudes for these 3 planets, we preferred to postpone the final
analysis and accumulate more measurements during the next 2
seasons. Extending by such amount the span and number of measurements
has permitted to confirm and check the robustness of the preliminary
solution.

\begin{figure}[t!]
\centering
\includegraphics[width=0.9\hsize]{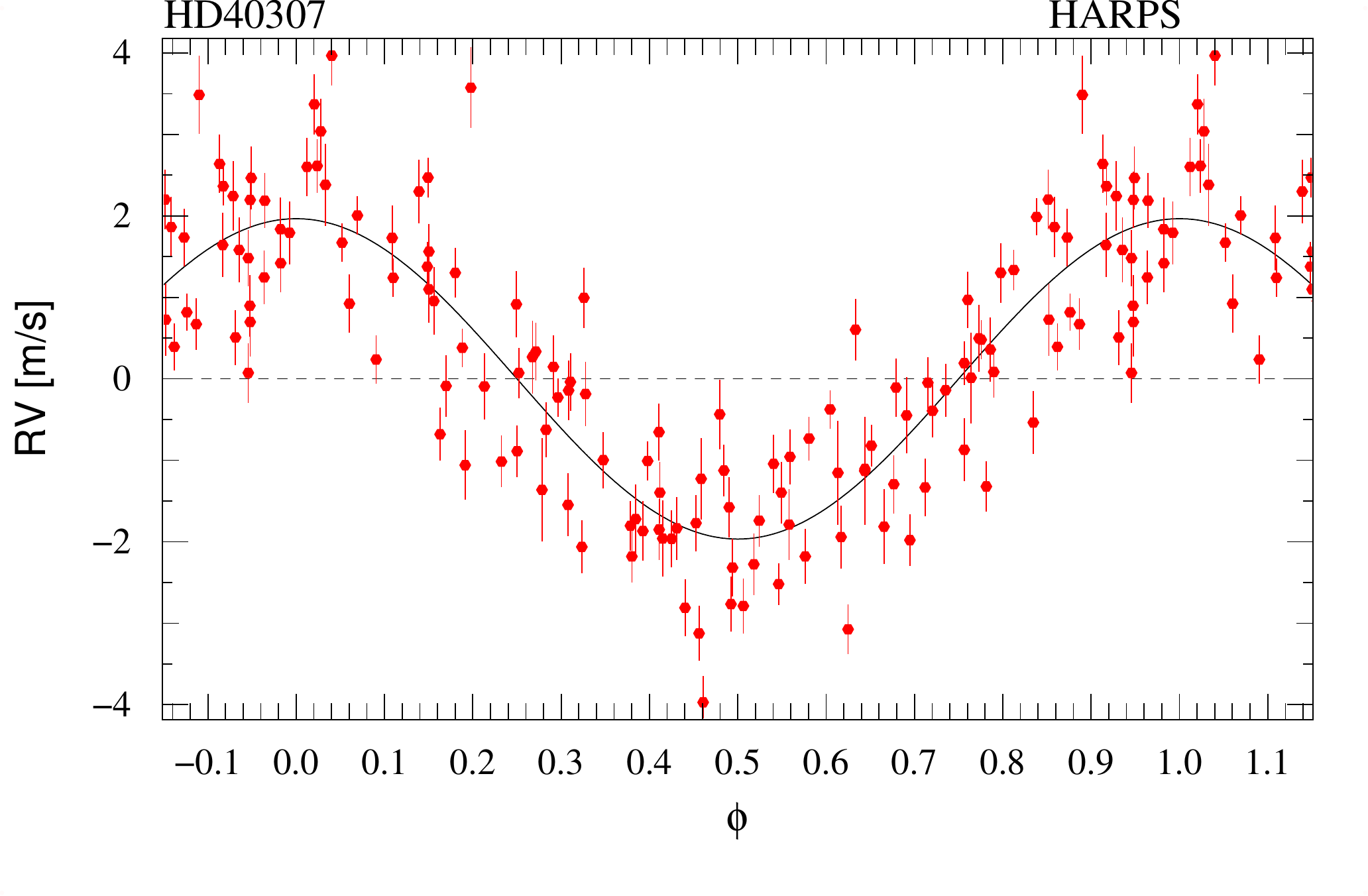}
\includegraphics[width=0.9\hsize]{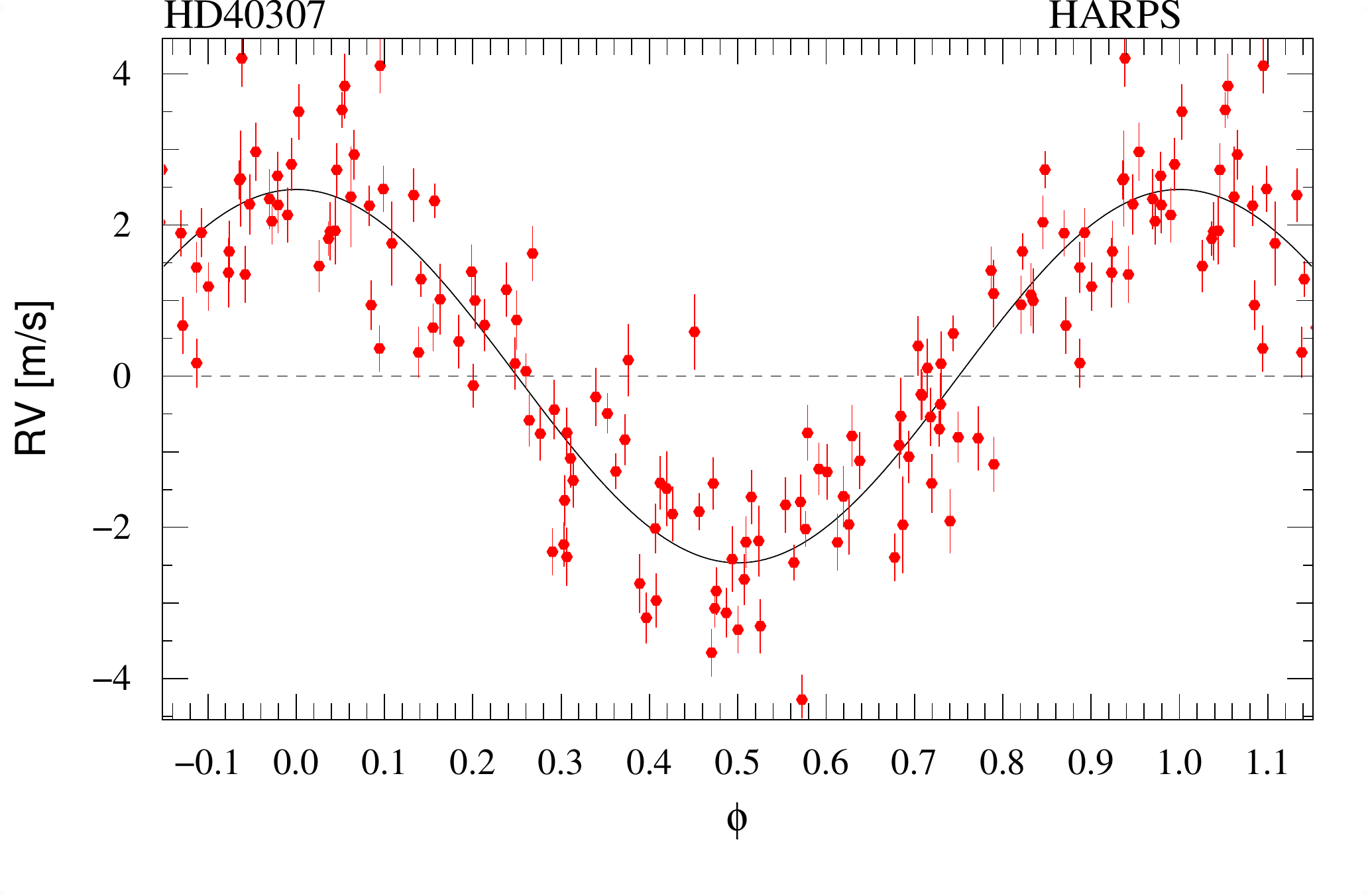}
\includegraphics[width=0.9\hsize]{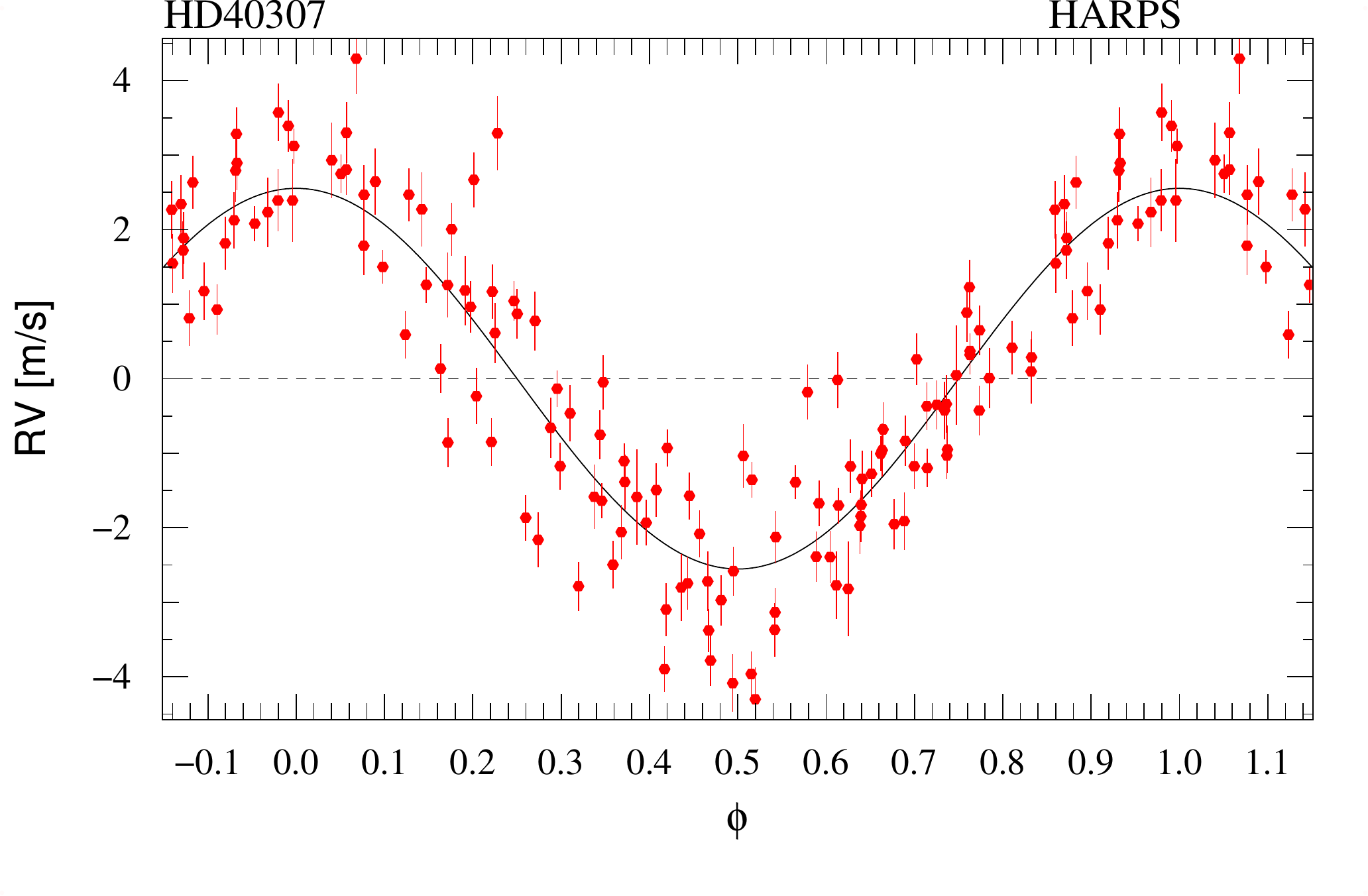}
\caption{Phase-folded radial velocities and Keplerian curve for 
each of the planets, after correction of the effect of the 2 other 
planets and of the drift.Curves related to planets b,c and d are 
illustrated from top to bottom, respectively.}
\label{mayorfig2}
\end{figure}

Using the now available data, three peaks are clearly identified in
the Fourier spectrum of the velocity measurements, they correspond to
periods of 4.2, 9.6 and 20.5 days (Fig.\,\ref{mayorfig1}). Fitting three
Keplerians to the radial velocity measurements, we obtain a solution
close to the one listed in Table\,\ref{TabOrb}, with periods
corresponding to the 3 mentioned peaks. Note here that the
conventional approach of iteratively determining the solution for the
different planets one after the other is very difficult because of the
short periods at play, moreover close to resonances. To find the final
solution we rather followed a genetic-algorithm approach to blindly
probe the complete parameter space.

\begin{figure}[t!]
\centering
\includegraphics[width=0.9\hsize]{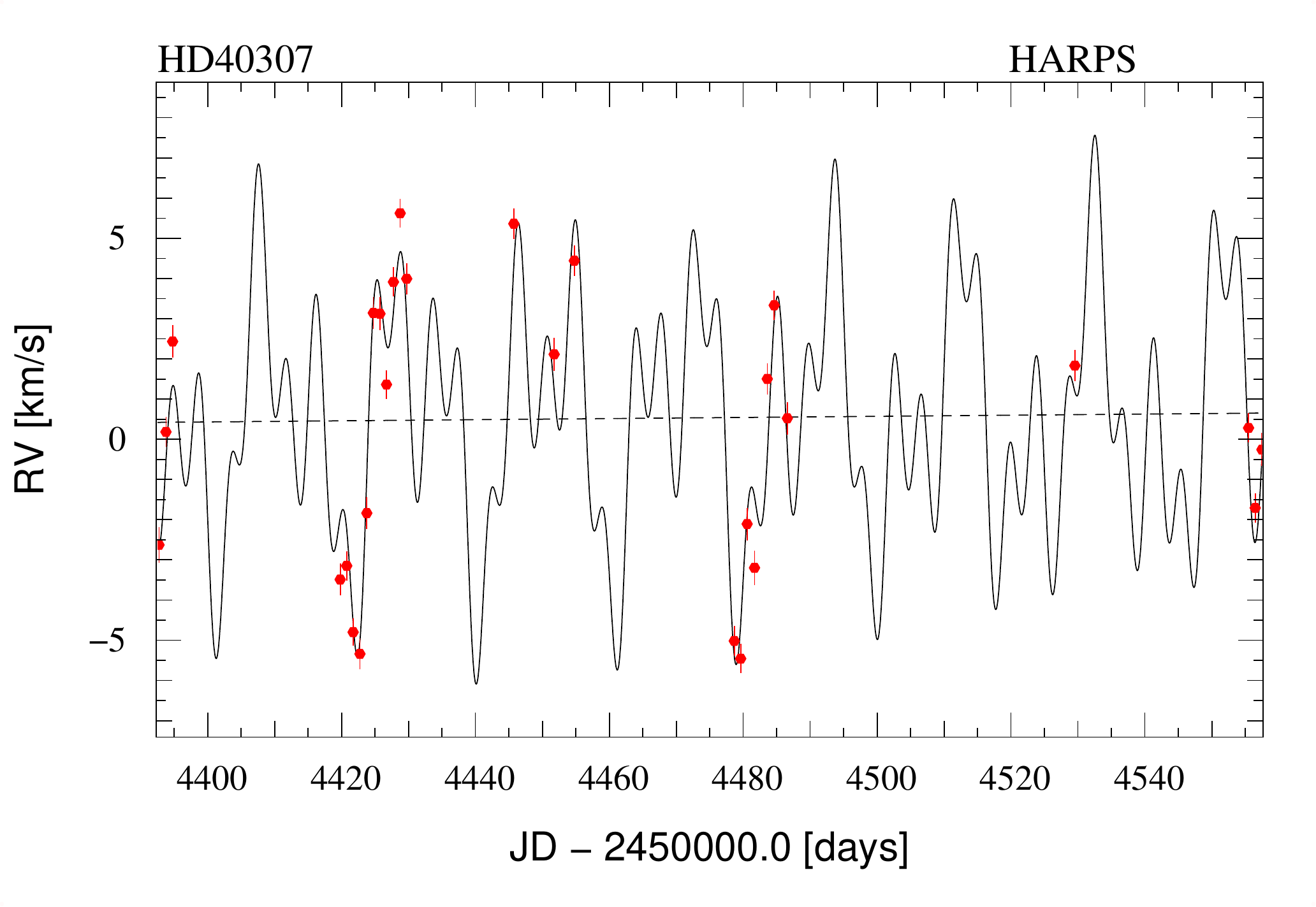}
\caption{Time window of HARPS relative radial velocities of HD\,40307 with the
3-keplerian model superimposed.}
\label{mayorfig3}
\end{figure}

\begin{figure}[t!]
\centering
\includegraphics[width=1.2\hsize]{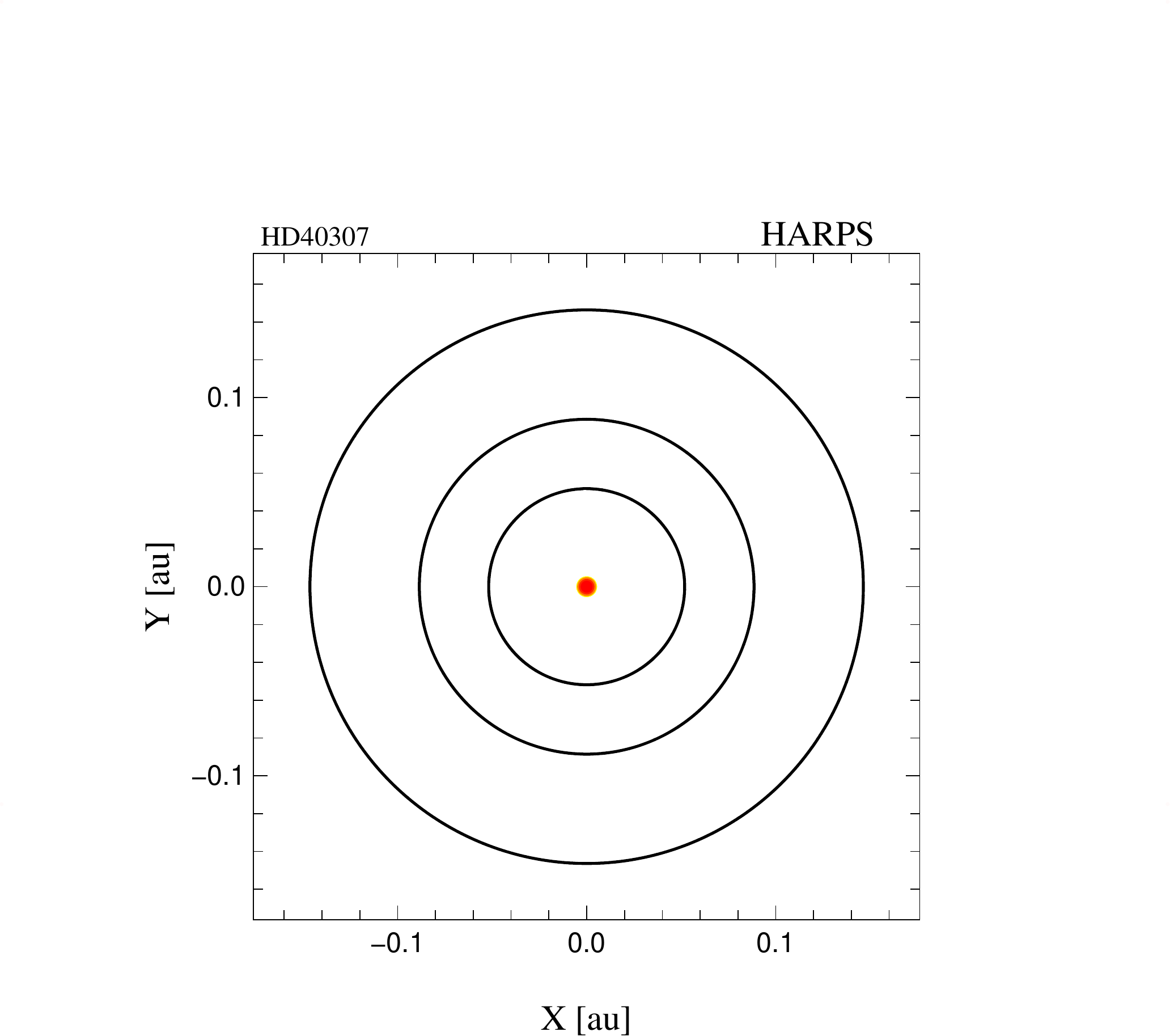}
\caption{Pole-on view of the 3 circular planetary orbits in the center
  of the HD\,40307 system.}
\label{mayorfig4}
\end{figure}

In a first fit the orbital eccentricities were let free. The derived
solution showed respectively non-significant eccentricities of $0.008
\pm 0.065$, $0.033 \pm 0.052$ and $0.037 \pm 0.052$ for the 3
planets. The residuals of the observed velocities around the best
3-planet solution is 0.94\,ms$^{-1}$. With circular orbits, the
residuals stay unchanged as well as the other orbital
parameters. The solution with 3 circular orbits is then the 
preferred one.  Additionally, the residuals (O-C) also exhibit a well 
defined linear drift over the span of our measurements. The best solution 
is finally  obtained by fitting  3 Keplerian circular orbits  plus a linear
drift to our 128 better measurements. The phase distribution of the radial 
velocities for the 3 planets are shown on Fig.\,\ref{mayorfig2}, a temporal 
window of the solution in Fig.\,\ref{mayorfig3}, and the top view 
of the system is illustrated in Fig.\,\ref{mayorfig4}.

The rms of the residuals to the fitted orbits (and drift), sigma(O-C)
is 0.85 ms$^{-1}$, still in excess with regards to the mean uncertainty of our 
observations.
Part of the extra noise could result from granulation, jitter related to the 
low level of chromospheric activity or instrumental effects. But we also 
have to consider the possibility of additional planets. Several potential
additional signals are present in the Fourier spectrum. Although they are 
non-significant, we have nevertheless tried a blind search for  4 Keplerians
plus a drift, without finding a convincing case for a fourth planet. 
Additional measurements are foreseen on the coming seasons to explore 
such a possibility and also to gain more insight on the companion responsible
of the observed linear drift.

The 3-planet solution is robust and precise. It is worth noticing the
rather low amplitudes of the 3 reflex motions induced by these planets
(1.97, 2.47 and 2.55\,ms$^{-1}$, respectively). Due to the rather large
number of measurements, all three amplitudes are rather precisely
determined.  The masses of these super-Earths are determined with a
precision of the order of 6\,\%  or better.

The ratios of periods $P_2/P_1 = 2.23$ and $P_3/P_2 = 2.13$ are
probably sufficiently distant to a rational value to exclude
resonances.

\begin{table*}[t!]
\caption{Fitted orbital solution for the planetary system around
  HD\,40307: 3 Keplerians plus a linear drift. To better estimate
  uncertainties on the adjusted parameters, the adopted solution has
  fixed circular orbits, the derived eccentricities being
  non-significant (see text).  }
\label{TabOrb}
\centering
\begin{tabular}{l l l c c c}
\hline\hline
\multicolumn{2}{l}{\bf Parameter} &\hspace*{2mm} 
& \bf HD\,40307\,b  & \bf HD\,40307\,c & \bf HD\,40307\,d \\
\hline
$P$ & [days] & & 4.3115 $\pm$ 0.0006 & 9.620 $\pm$ 0.002 & 20.46 $\pm$ 0.01 \\
$T$ & [JD-2400000] & & 54562.77 $\pm$ 0.08 & 54551.53 $\pm$ 0.15 
& 54532.42 $\pm$ 0.29 \\
$e$ & & & 0.0 & 0.0 & 0.0   \\
$\omega$ & [deg] & & 0.0 & 0.0 & 0.0 \\
$K$ & [m s$^{-1}$] & & 1.97 $\pm$ 0.11 & 2.47 $\pm$ 0.11 & 4.55 $\pm$ 0.12 \\
$V$ & [km s$^{-1}$] & & \multicolumn{3}{c}{31.332}  \\
$drift$ & [m s$^{-1}$/yr] & & \multicolumn{3}{c}{0.51 $\pm$ 0.10}  \\
$f(m)$ & [10$^{-14} M_{\odot}$] & & 0.35 & 1.53 & 3.59 \\
$m_2 \sin{i}$ & [$M_{\oplus}$] & & 4.2 & 6.8 & 9.2 \\
$a$ & [AU] & & 0.047 & 0.081 & 0.134 \\
\hline
$N_{\mathrm{meas}}$ & & & \multicolumn{3}{c}{135} \\
{\it Span} & [days] & & \multicolumn{3}{c}{1628} \\
$\sigma$ (O-C) & [ms$^{-1}$] & & \multicolumn{3}{c}{0.85} \\
$\chi^2_{\rm red}$ & & & \multicolumn{3}{c}{2.57} \\
\hline
\end{tabular}
\end{table*}

\section{Summary and discussion}

We report the detection of 3 super-Earth planets orbiting the low
metallicity K dwarf HD\,40307, a star located at only 13\,pc from the
Sun. The high precision radial velocities acquired with the HARPS
spectrograph on the ESO 3.6-m telescope enabled this discovery. The
3-Keplerian fit reveals the presence of 3 low-mass planets. The
closest one HD\,40307\,b with $m_2\sin{i}$\,=\,4.2\,M$_{\oplus}$ is
presently the lightest exoplanet detected around a main sequence
star. The two other planets, with masses of 6.9 and 9.2\,M$_{\oplus}$
also belong to the category of super-Earths. All the 3 planets are on
circular orbits.  It is amazing to notice that the global rms of the
135 measurements, before fitting any planets, was only
2.94\,ms$^{-1}$.  The sigma(O-C) after the 3-planet Keplerian fit and drift is
0.85\,ms$^{-1}$.  We will continue the velocity monitoring to better
characterize the longer period 4th object bound to the system,
revealed by the additional observed linear drift of the radial velocities.

Available Spitzer IRS data of HD\,40307 do not show any IR excess
in the 10--40\,$\mu$m region of the spectra (Augereau, private communication). 
No warm dust disk is thus detected in the inner regions of the system, 
unlike the case of HD\,69830, the star harbouring a trio of Neptune planets 
\citep{Lovis-2006} and for which an observed IR excess indicates the presence 
of a debris disk, possibly under the form of an asteroid belt 
\citep{Beichman-2005}.
 
The characterization of multi-planetary systems with very low-mass
planets require a rather large number of measurements.  After 4.5
years of the HARPS programme and with the improved reduction 
software, several dozens of planets with masses less than 30 Earth-masses 
and period less than 50 days have been detected. Coming observations will
confirm these detections and allow us to fully characterize the
systems.  The domain of Neptune-type and rocky planets will be
drastically boosted in a near future with these detections.  In
particular we expect to have enough systems to revisit the emerging
properties for these low mass planets as tentatively discussed by
\citet{Mayor-2008}:\\
\indent
- is the mass-distribution of exoplanets bimodal?  \\
\indent
- Is the correlation between host star metallicity and occurrence
frequency of Neptune-type planets (or smaller) still existing?\\
\indent
- What is the frequency of multiplanetary systems with these low mass
planets?\\
\indent
- What is the frequency of Neptune or rocky planets orbiting G and K
dwarfs?  A first estimate based on the HARPS high-precision survey
suggests a frequency of 30\,$\pm$\,10\,\% in the narrow range of
periods shorter than 50 days.
  

One of the most exciting possibility offered by this large emerging
population of low-mass planets with short orbital periods is the
related high probability to have transiting super-Earths among the
candidates. If detected and targeted for complementary observations,
these transiting super-Earths would bring a tremendous contribution to
the study of the expected diversity of the structure of low-mass
planets.

\begin{acknowledgements}
  The authors thank the different observers from the other HARPS GTO
  sub-programmes who have also measured HD\,40307. We also thank
  J.-C. Augereau, E. Difolco, and J. Olofsson for providing their reduction 
  of the Spitzer IRS data on HD\,40307.  We finally are grateful to
  the Swiss National Science Foundation for its continuous interest
  supporting the project.
\end{acknowledgements}

\bibliographystyle{aa}
\bibliography{udry_articles}

\end{document}